# Prototipo de video juego activo basado en una cámara 3D para motivar la actividad física en niños y adultos mayores

Benjamín Ojeda Magaña, José G. Robledo Hernández, Leopoldo Gómez Barba, Víctor M. Rangel Cobián

*Universidad de Guadalajara, Departamento de Ingeniería de Proyectos, CUCEI

Resumen

El presente documento describe cómo se desarrolló un prototipo de videojuego para fomentar la actividad física en niños y adultos mayores. El prototipo está integrado por una computadora portátil, una cámara con sensores 3D y opcionalmente requiere una pantalla LCD o un proyector.

El componente de programación de este prototipo fue desarrollado en Scratch que es un lenguaje de programación orientado a niños, por lo que las posibilidades de desarrollar un juego al gusto de los usuarios son muy reales.

La idea de crear un prototipo de este tipo se origina en la idea de aportar una opción que promueva la actividad física en niños y adultos, ya que la falta de ejercicio físico es un factor preponderante en el desarrollo de enfermedades crónico-degenerativas como son la diabetes y la hipertensión por mencionar las de mayor incidencia.

Como resultado de esta propuesta se logró desarrollar un prototipo de videojuego activo de prueba basado en un juego de ping-pong el cual permite la interacción de niños y adultos de una manera divertida y al mismo tiempo se fomenta la realización de actividades físicas que pueden impactar de manera positiva en la salud de los usuarios.

## 1. Introducción

El concepto denominado actividad física se refiere a cualquier tipo de movimiento corporal producido a través de los músculos y que requiere un gasto de energía. La actividad física puede tener muchas variantes, como, por ejemplo: nadar, jugar, saltar, correr y montar en bicicleta entre otras. La actividad o ejercicio físico tiene múltiples beneficios para personas de todas las edades desde niños, adolescentes, jóvenes y adultos en general. La actividad física puede ayudar a tener un buen estado de salud general, ayuda a prevenir el sobrepeso y la obesidad, los cuales son factores muy importantes por considerar en la prevención de enfermedades como la diabetes, la hipertensión y el cáncer. Adicionalmente, la actividad física es una de varias herramientas que pueden favorecer la relajación y la descarga de tensión que a su vez contribuyen a mejorar la autoestima.

Existen diversas estrategias para promover en las personas el desarrollo de actividad física en casa, tales como, realizar tablas de ejercicios propuestos por especialistas, inscribirse a clases de deportes por internet, bailar al ritmo de la música, utilizar bicicletas estáticas, realizar estiramientos, caminar a través de las habitaciones y algunas otras más(Carlemany,

2022). En el caso del presente trabajo, se presenta un prototipo que se ha desarrollado bajo el enfoque de una estrategia también utilizada para promover la actividad física y se enmarca en el concepto denominado Video Juegos Activos (AVG o Active Video Game por sus siglas en idioma inglés).

Algunas definiciones con relación a lo que este prototipo es y representa se muestran a continuación:

"Un juego es una actividad interactiva voluntaria, en la que uno o más jugadores siguen reglas que restringen su comportamiento, difundiendo un conflicto artificial que termina en un resultado cuantificable" (Zimmerman, 2004).

"Un videojuego es un juego al que jugamos gracias a un aparato audiovisual y que puede estar basado en una historia" (Esposito, 2005).

Un video juego activo es "un videojuego que requiere actividad física más allá de la de un juego pasivo (es decir, juegos portátiles convencionales). Los videojuegos activos se basan en tecnología que rastrea el movimiento o la reacción del cuerpo para que el juego progrese." (LeBlanc et al., 2013).

Realizando una revisión a la literatura en relación con el uso de video juegos activos para promover la actividad física en personas se encontraron resultados variables. Se tiene por ejemplo a Biddiss (Biddiss & Irwin, 2010) que realizó una revisión sistemática de literatura que abarca un período comprendido entre 1998 y 2010 encontrando que el uso de Video Juegos Activos (AVG) elevó el gasto energía en jóvenes en hasta un 222%, sin embargo, también encontró una alta variabilidad en el gasto de energía dependiendo de si el juego implica el uso de extremidades superiores o inferiores. En otro estudio posterior (LeBlanc et al., 2013) se encontró que ciertos estudios controlados mostraron que los AVG aumentan de forma aguda la actividad física de intensidad ligera a moderada; sin embargo, los hallazgos sobre si los AVG conducen a aumentos en la actividad física habitual o a disminuciones en el comportamiento sedentario, y cómo lo hacen, no es claro. Aunque los AVG pueden generar algunos beneficios para la salud en poblaciones especiales, los autores indican que no encontraron pruebas suficientes para recomendar los AVG como un medio para incrementar la actividad física diaria. Por su parte (Foley & Maddison, 2010) encontraron que en comparación con los videojuegos tradicionales no activos, los videojuegos activos provocaron un mayor gasto de energía, que fue similar en intensidad a la actividad física de intensidad leve a moderada. Los estudios de intervención indican que los videojuegos activos pueden tener el potencial de aumentar la actividad física de vida libre y mejorar la composición corporal en los niños. Estudios más recientes como los de (Gómez et al., 2019)han encontrado que los videojuegos activos y la gamificación pueden utilizarse en programas educativos para aumentar la motivación de los niños y adolescentes en el ejercicio físico. Estos últimos autores consideran que la profesión de enfermería puede jugar un papel fundamental en la educación para la salud. Adicionalmente consideran que, los programas educativos para promover el ejercicio físico y los hábitos saludables deben diseñarse y desarrollarse desde los Centros de Atención Primaria de Salud y, también, en las escuelas. Y es que, la introducción de la gamificación en estos programas educativos puede mejorar la calidad de vida de los niños que padecen obesidad infantil gracias a la adquisición de hábitos saludables. Por otro lado (Williams & Ayres, 2020) consideran que el resultado indica que los juegos de exergaming o los videojuegos activos pueden ser una herramienta eficaz para mejorar la actividad física en los adolescentes que será más aceptable y sostenible que muchos enfoques convencionales.

Los resultados encontrados por los investigadores en general muestran que, aunque de manera variable en general la estrategia de video juego activo es preferible a la de los video juegos convencionales, por el simple hecho de que estos sí promueven conductas no sedentarias en las personas.

## 2. Características del prototipo de videojuego activo.

Como se mencionó anteriormente, el objetivo de este prototipo es el de promover que el usuario realice actividad física y a la vez lo haga de una manera divertida. Teniendo en cuenta este objetivo, se diseñó un prototipo de videojuego activo basado en las prestaciones que un sensor o una cámara 3D nos pueden proporcionar como elementos de detección de movimientos a distancia, un equipo de cómputo para procesar el programa diseñado para el videojuego y una pantalla o proyector que muestre las secuencias gráficas que harán que el usuario o jugador se involucre realizando la actividad física.

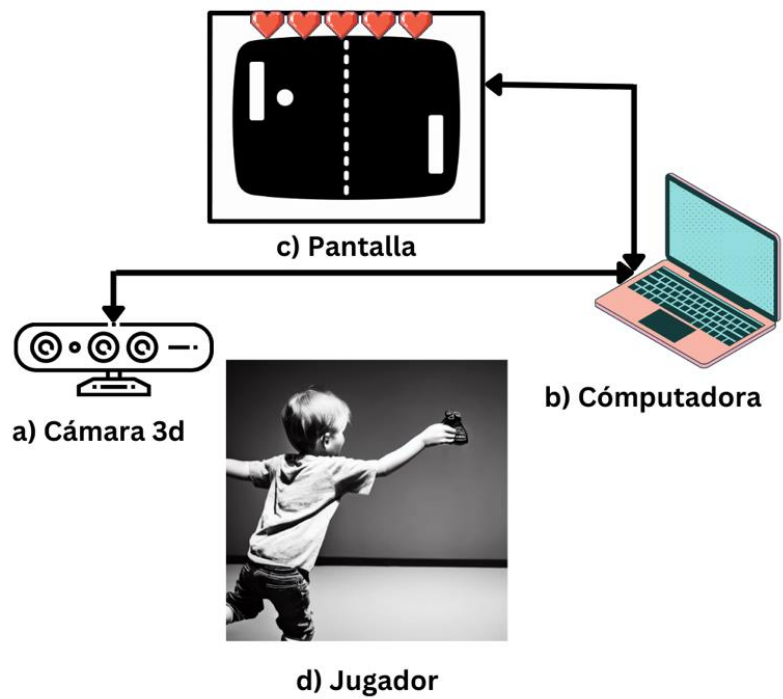

*Figura 1* *Esquema que muestra los componentes del prototipo de video juego activo: a) Cámara 3D, b) Computadora portátil, c) Pantalla y d) Jugador.*

En la *Figura 1* se muestran los componentes principales del sistema los cuales son:

a) Cámara o sensor 3D.

La cámara 3D es un sensor cuya función es detectar todos los movimientos del jugador a cierta distancia. Es un componente fundamental ya que tiene que ver con la interacción entre el jugador y el programa del videojuego, es decir, la secuencia del video juego en su mayor parte depende del curso que le dan los movimientos del usuario.

Los sensores de profundidad 3D están basados en un fundamento en el cual para cada pixel de la imagen capturada se determina la distancia que hay entre el sensor 3D y el objeto de interés, en este caso el jugador, mediante la medición de un tiempo de retardo. Actualmente, debido a los avances tecnológicos, cada vez más se puede tener acceso a cámaras 3D de bajo costo las cuales permiten la captura de imágenes en tiempo real con la información adicional de distancia asociada a los objetos de interés. En el caso del sensor 3D utilizado para este prototipo es del tipo Kinect Versión 1. Las características básicas (Del Valle-Hernandez, n.d.) de este sensor según el sitio web *programarfacil* [1] son las siguientes:

- Cámara de vídeo de color RGB. Funciona a modo de webcam, capturando las imágenes en vídeo. El sensor Kinect utiliza esta información para obtener detalles sobre objetos y personas en la habitación.
- Emisor IR. El emisor de infrarrojos es capaz de proyectar una luz infrarroja en una habitación. Según la luz infrarroja incide sobre una superficie, el patrón se distorsiona. Esta distorsión es leída gracias a su otro componente, una cámara de profundidad.
- Cámara de profundidad. Analiza los patrones infrarrojos emitidos por el emisor y es capaz de construir un mapa 3D de la habitación y de todos los objetos y personas que se encuentran dentro de ella.
- Conjunto de micrófonos. El sensor Kinect tiene incorporado cuatro micrófonos de precisión capaces de determinar de dónde vienen los sonidos y las voces. También es capaz de filtrar el ruido de fondo.
- Motor de inclinación. Este motor tiene la capacidad de ajustar sobre la base, el sensor Kinect. Es capaz de detectar el tamaño de la persona que está delante, para ajustarse arriba y abajo según convenga.

b) Computador y plataforma de programación.

El computador es el encargado de almacenar y ejecutar el programa del video juego. Inicialmente, el programa del videojuego recibe las condiciones iniciales de la partida, como son, el número de jugadores (uno o dos), el grado de dificultad de las acciones físicas que se van a exigir que normalmente tienen que ver con la rapidez de respuesta, la selección de las partes del cuerpo con las cuales se va a interactuar (cabeza o mano) y finalmente con la presentación del puntaje logrado (score) ya que se trata de un juego. El tipo de computadora utilizado para este prototipo en específico ha sido una computadora portátil con procesador Intel Core i7 y memoria RAM de 16 GB. Sin embargo, se debe mencionar que se pueden utilizar equipos con menor cantidad de recursos, incluso es posible utilizar computadoras compactas o microcomputadoras de tipo raspberry pi. Respecto al lenguaje de programación bajo el cual se desarrolló este prototipo se mencionó anteriormente que se denomina Scratch. Debe mencionarse que Scratch no es solamente un lenguaje de programación, sino que es también una comunidad para promover la programación entre niñas y niños con alcance a nivel mundial.

---

[1] https://programarfacil.com/podcast/86-sensor-kinect-inteligencia-artificial/

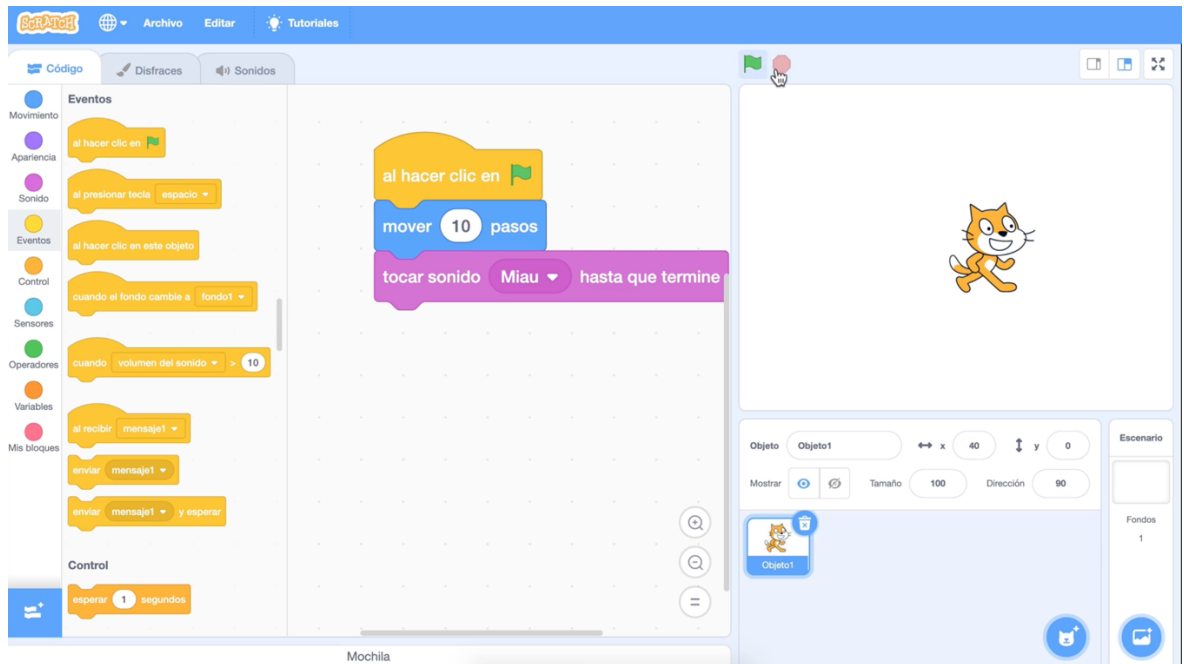

*Figura 2* Interfaz gráfica de desarrollo de la plataforma de programación denominada Scratch.

En la Figura 2 se muestra la interfaz gráfica de la plataforma de desarrollo de Scratch que puede funcionar directamente desde la web. Esta plataforma facilita la creación de historias, juegos y animaciones, a través del uso de bloques de programación que controlan la ejecución de acciones como movimiento, apariencia, sonido, eventos, control de iteraciones, control de sensores no solo como el ratón y el teclado sino también como el Kinect y su cámara 3D. Desde luego también se dispone de bloque para manejar variables y operadores.

c) Pantalla o Proyector.

La pantalla o proyector tienen la función de presentar las secuencias visuales propias del desarrollo del videojuego las cuales estarán asociadas o vinculadas a través del programa en ejecución y a la detección del sensor 3D, es decir, dichas secuencias se presentarán en razón a la actividad física que el jugador desarrolle. En el desarrollo de este prototipo se han probado pantallas planas de diferentes dimensiones y también se ha utilizado proyectores para mostrarle imagen en los muros de alguna habitación. dependiendo de la calidad que se quiera obtener debe ser el tipo de pantalla a utilizar. Se pueden utilizar pantallas de bajo costo y de mediana dimensión las cuales cumplen con los requisitos básicos que debe tener este prototipo.

   d) Jugador.

Es el elemento central y motivo por el cual se desarrolló este prototipo. De su actividad física desarrollada depende todo el demás funcionamiento del prototipo.

## 3. Método implementado en el prototipo de videojuego activo

El método implementado para la ejecución del prototipo de videojuego activo se muestra en la figura 3.

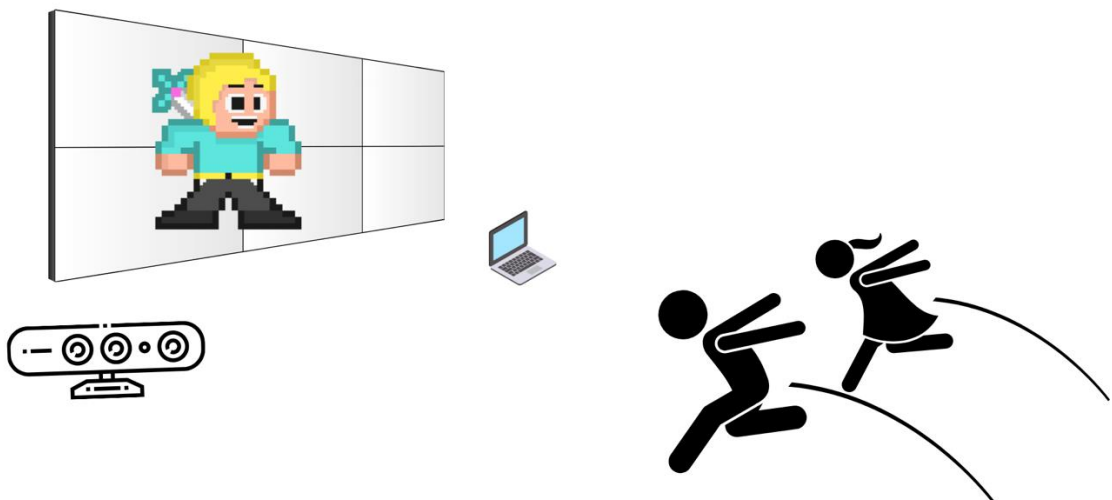

*Figura 3* *Disposición de loe elementos del prototipo de videojuego activo durante su ejecución.*

En la figura 3 se muestra el orden y disposición que deben mantener los elementos componentes del prototipo de videojuego activo durante su ejecución. El orden de ejecución es el siguiente:

   I.   Los jugadores deben colocarse de forma frontal y viendo hacia dos de los componentes del prototipo que son el sensor Kinect y la pantalla.
   II.  Encender los componentes en este orden: a) Pantalla b) Computadora y c) Kinect.
   III. El control de la interface se realiza a distancia a través del Kinect, haciendo un gesto específico.
   IV.  Se selecciona el número de jugadores.
   V.   Se selecciona el grado de dificultad.
   VI.  Se selecciona la parte del cuerpo que se usará durante el juego.
   VII. Se inicia el juego.

VIII. Se desarrolla el juego hasta que alguien gane.
IX. Se observa el marcador y se decide si se termina o continúa otro partido.

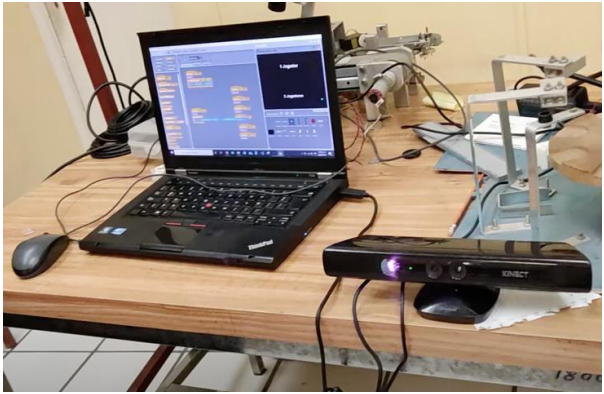 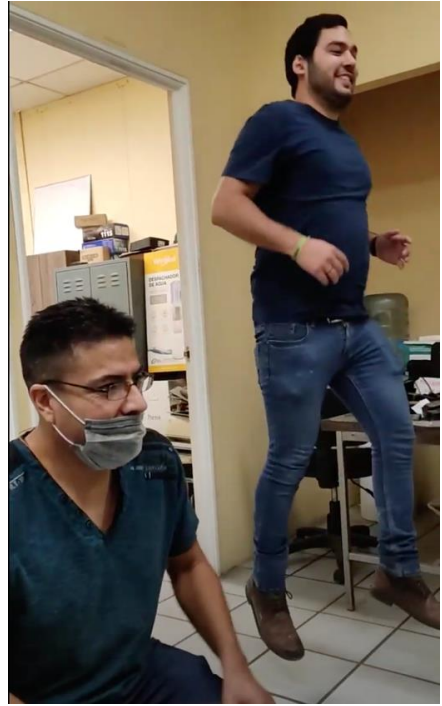

a                                            b

*Figura 4* a) Fotografía de los componentes del prototipo durante una prueba del desarrollo y b) usuarios tomando parte en el juego.

En la figura 4 se incluye un par de fotografías que muestran parte de las pruebas del desarrollo del prototipo en el laboratorio, así como la puesta en marcha y prueba por parte de un par de jugadores adultos.

## 4. Funciones y bloques de código desarrollado para un juego de Ping Pong

El juego de prueba desarrollado para el presente prototipo consiste en un juego de ping-pong que se juega ya sea con la cabeza o con la mano y que dispone de dos niveles de dificultad (fácil y difícil). La puesta en marcha del juego se realiza a través de movimientos de las manos que son detectados por los sensores del Kinect y así se inicia la selección de las condiciones iniciales de la partida.

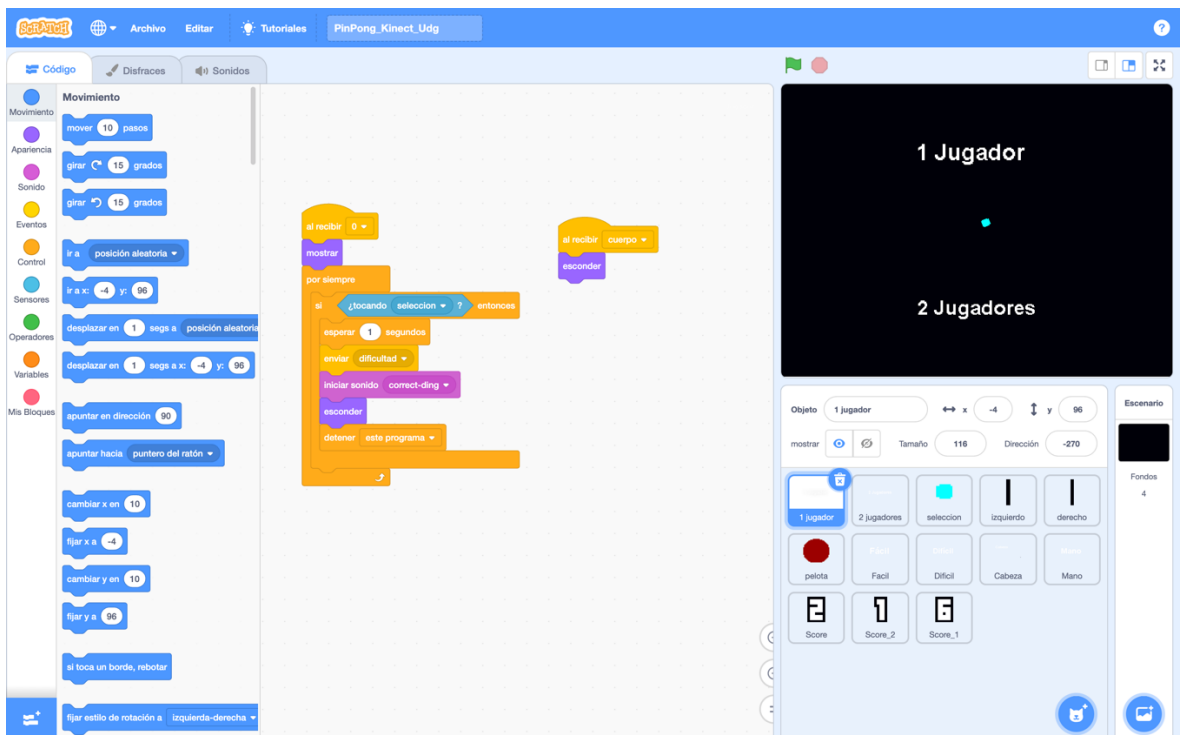

***Figura 5*** *Pantalla que muestra parte del desarrollo del juego de prueba ping-pong para ser ejecutado en el prototipo de videojuego activo.*

En la figura 5 se muestra una captura de pantalla referente al desarrollo de un videojuego de prueba, específicamente se trata de una réplica de un juego de ping-pong que fue popular en la década de los 80's. Este video juego ha sido adaptado para ejecutarse como un videojuego activo, ya que los elementos que sustituyen a las palancas (joysticks) de la consola de videojuegos son dos partes del cuerpo que se pueden elegir, ya sean las manos para un menor grado de dificultad o bien la cabeza, lo cual incrementa el grado de dificultad. Estas partes del cuerpo funcionan como si fueran las paletas o barreras con las cuales se tiene que devolver la pelota. En la figura 5 también se muestran diferentes objetos o módulos que se tuvieron que programar como parte del videojuego. Entre los módulos desarrollados se encuentran:

1) Un jugador.
2) Dos jugadores.
3) Selección.
4) Izquierdo.
5) Derecho.
6) Pelota.
7) Fácil.
8) Difícil.
9) Cabeza.
10) Mano
11) Score

Cada módulo está desarrollado en bloques de programación de Scratch. Algunos son muy fáciles de entender debido a lo intuitivo y visual que es la plataforma, y otros son más extensos y requieren de mayor atención para su comprensión. A continuación, se presentan los códigos de los objetos enumerados anteriormente.

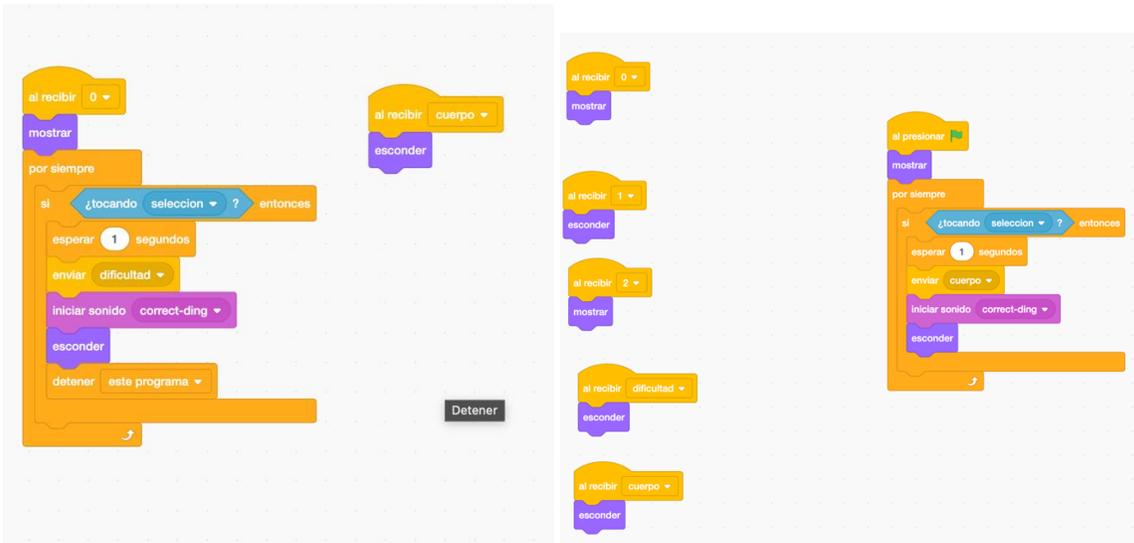

*Figura 6 Módulos para la selección de uno o dos jugadores.*

Los módulos para la selección de uno o 2 jugadores inicializan el juego de la siguiente manera. Si se selecciona un solo jugador, el adversario será la computadora propiamente. en caso de seleccionar la opción 2 jugadores, el videojuego se llevará a cabo entre 2 usuarios reales. La figura 6 muestra los bloques de código utilizadas para controlar la selección de cantidad de jugadores.

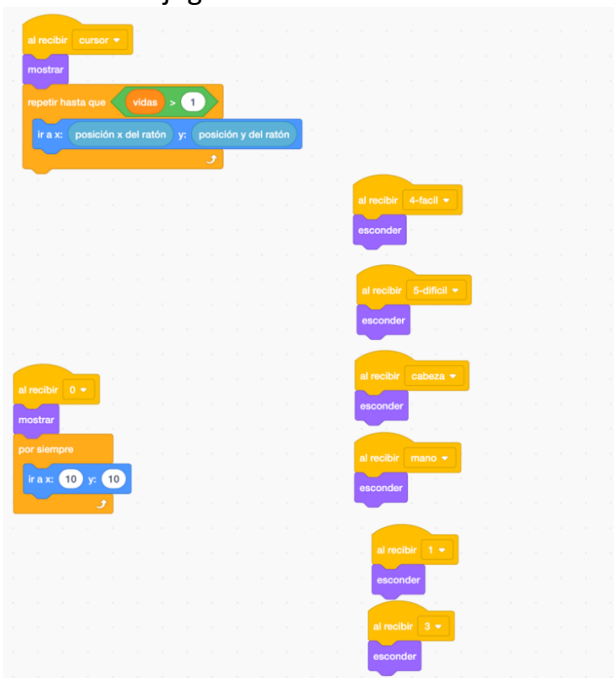

*Figura 7 Módulo de control de selección.*

La figura 7 muestra el módulo de selección. Este módulo se encarga de controlar la ubicación del cursor a través de los movimientos que son reconocidos por los sensores del Kinect. Asimismo, redirecciona el programa según las opciones señaladas.

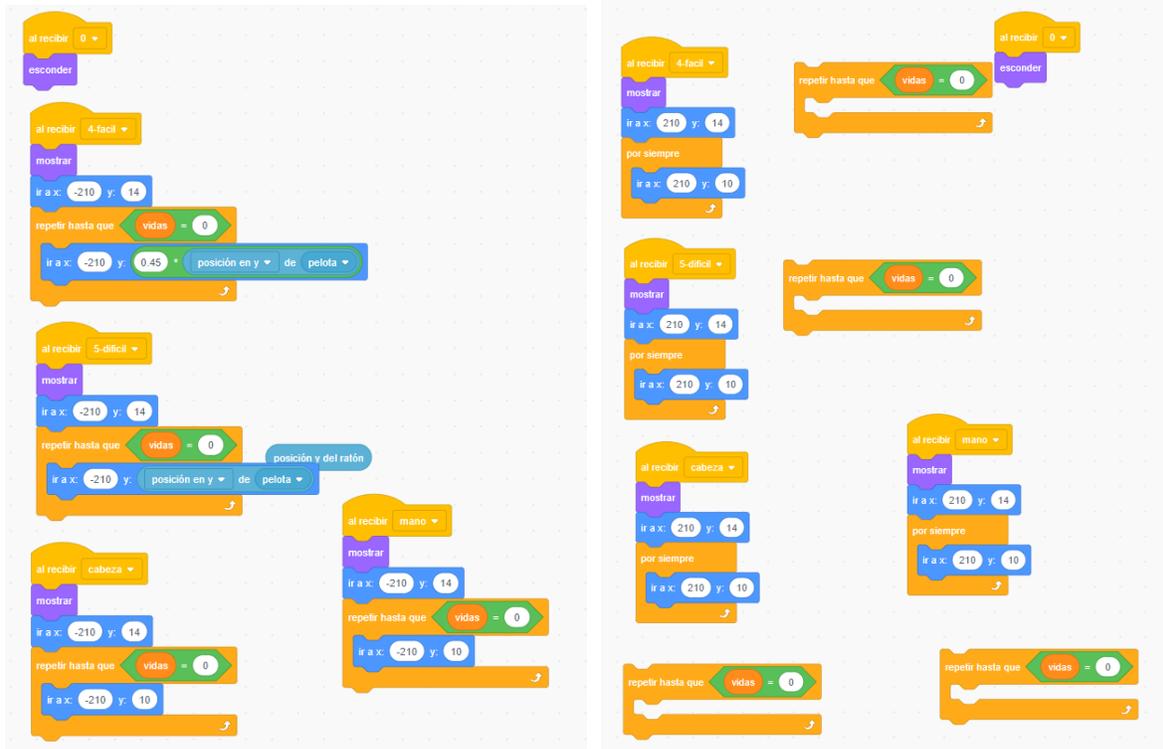

*Figura 8* Módulos Izquierdo y Derecho.

En la figura 8 muestra un par de módulos denominados izquierdo y derecho. Estos 2 módulos se encargan de mostrar y controlar las barras o paletas que chocan con la pelota de ping-pong. Dependiendo de la parte del cuerpo utilizada y del grado de dificultad seleccionado será la trayectoria y velocidad que se le imprime a la pelota.

En la figura 9 se muestran otro par de módulos cuya función es seleccionar las opciones de grado de dificultad, ya sea nivel fácil o nivel difícil. este par de módulos interacciona a través del sensor Kinect para que a través de la utilización de las manos se puede controlar el cursor y se pueda seleccionar las opciones deseadas.

Asimismo, en la figura 10, se muestran 2 módulos que se denominan mano y cabeza. Estos módulos tienen la función de seleccionar la parte del cuerpo con la cual se jugará la partida. Dependiendo de la parte del cuerpo utilizada se controla de manera implícita un cierto nivel de dificultad adicional. De manera similar a otros módulos, estos dos interaccionan con el sensor de Kinect para poder manipular el cursor.

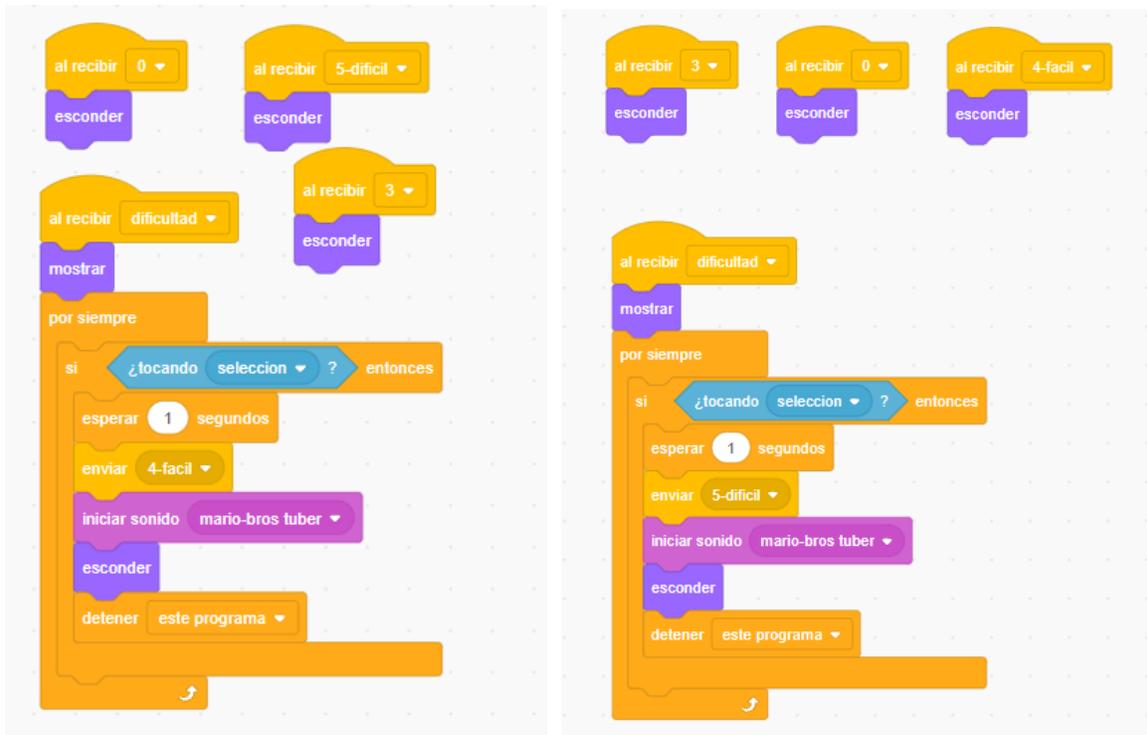
*Figura 9* Módulos para seleccionar el grado de dificultad.

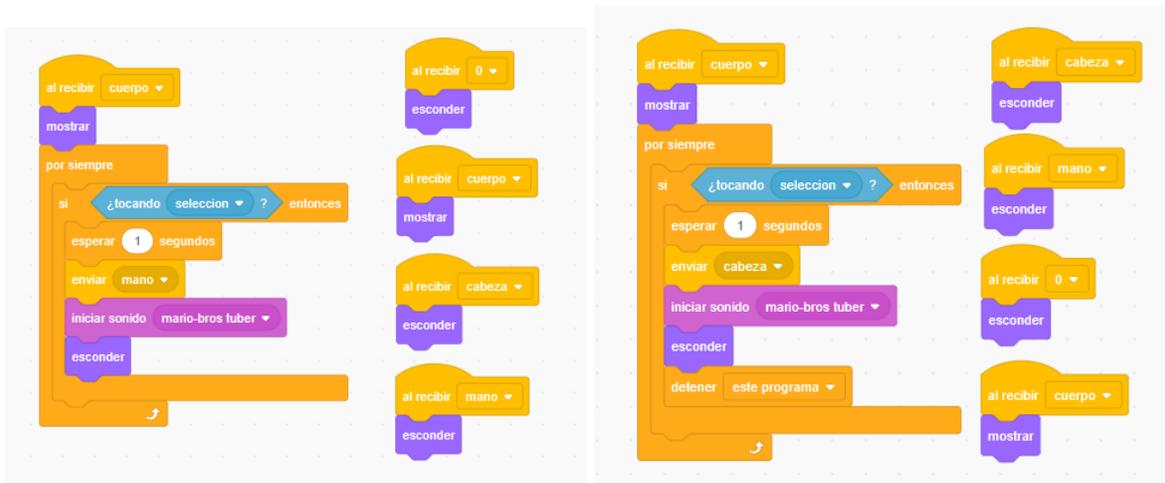
*Figura 10* Módulos para seleccionar la parte del cuerpo a utilizar en el juego.

En la figura 11 se muestra parte del código de la programación del módulo pelota. Este módulo es el más grande en cantidad de instrucciones y por lo mismo su tamaño no permite mostrarlo completo o de una manera que sea legible. Este módulo se encarga de controlar el comportamiento de la pelota. Para ello, toma en cuenta todos los parámetros seleccionados previamente como son los niveles de dificultad, la parte del cuerpo que se está utilizado y el número de jugadores. En respuesta a estos valores, el módulo asigna valores a ciertas variables como la velocidad, el puntaje y la cantidad de vidas.

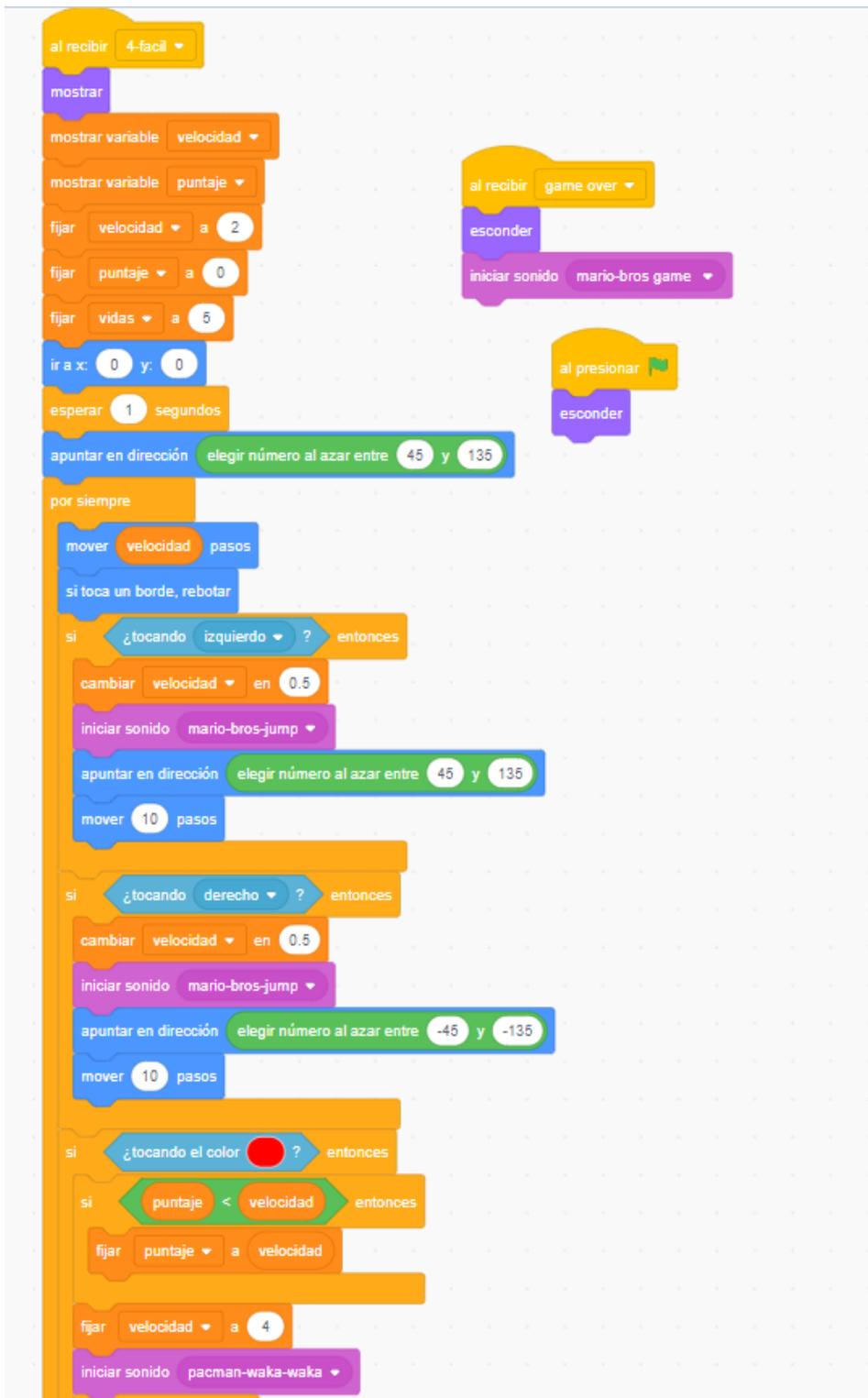

*Figura 11* Módulo de control del comportamiento de la pelota.

## 5. Conclusiones

En este trabajo se presenta el uso de una cámara 3D y de una plataforma de programación orientada a niños y también puede ser del gusto de los adultos. El prototipo de videojuego activo se desarrolló basándose en las dos consideraciones anteriores y como resultado se generó un videojuego de prueba basado en una réplica del juego de ping-pong popular en la década de los 80's. Este videojuego activo de prueba se ha probado con personas adultas y niños de edades entre 6 y 8 años y ha resultado interesante y aceptado. No obstante que se ha utilizado una plataforma de hardware que es antigua como lo es el Kinect, se tiene la posibilidad de migrar a dispositivos más actuales y de bajo costo para continuar el desarrollo y presentar otra versión más económica y que de esta manera, permita proponerse como otra opción que promueva la actividad física en niños y adultos con el fin principal de promover estados de bienestar en cuanto a salud se refiera.

## 6. Referencias